%% file: eprint.tex
\documentclass[10pt]{article}
\usepackage{graphicx}

\newcommand{\ttbar}{$\rm{t}\bar{\rm t}$}

\def\Title#1{\begin{center} {\Large #1 } \end{center}}
\def\Author#1{\begin{center}{ \sc #1} \end{center}}
\def\Address#1{\begin{center}{ \it #1} \end{center}}

\newcommand\pubblock{\rightline{\begin{tabular}{l} Proceedings of the Second Annual LHCP\\ \pubnumber\\
         \pubdate  \end{tabular}}}

\newenvironment{Abstract}{\begin{quotation} \begin{center} 
             \large ABSTRACT \end{center}\bigskip 
      \begin{center}\begin{large}}{\end{large}\end{center} \end{quotation}}

\newenvironment{Presented}{\begin{quotation} \begin{center} 
             PRESENTED AT\end{center}\bigskip 
      \begin{center}\begin{large}}{\end{large}\end{center} \end{quotation}}

\input econfmacros.tex

\usepackage{color}

\newcommand\pubnumber{ CMS-2014-204 }

\newcommand\pubdate{\today}

\def\affiliation{
for the ATLAS, CDF, CMS and D0 Collaborations} 

\begin{document}

\large
\begin{titlepage}
\pubblock

\vfill
\Title{  Top quark production  }
\vfill

\Author{ Mara Senghi Soares  }
\Address{\affiliation}
\vfill
\begin{Abstract}

We review the current status of the cross sections measurement of the top-quark at the LHC and at the Tevatron.  
Total production cross sections, 
studies using single top quark events and differential \ttbar~cross sections are discussed. The associated production of top quark pairs with photons, 
Z and W bosons,   including \ttbar Z and \ttbar W measurements shown for the first time at LHCP2014, are presented.

\end{Abstract}
\vfill

\begin{Presented}
The Second Annual Conference\\
 on Large Hadron Collider Physics \\
Columbia University, New York, U.S.A \\ 
June 2-7, 2014
\end{Presented}
\vfill
\end{titlepage}
\def\thefootnote{\fnsymbol{footnote}}
\setcounter{footnote}{0}
%

\normalsize 


\section{Introduction}

The physics of the top quark is closely related to some of the most interesting issues that still remain 
open in the Standard Model (SM).
Due to its large
mass, among all elementary particles the top quark has the largest couplings 
to other SM particles, e.g. the Higgs boson, and, potentially, to particles in new physics models.
Also, the strenght of the top quark coupling to the Higgs boson ($\approx1$) suggests that the top quark may play a special role in the mechanism of electroweak. 
Associated production of top pairs with the Higgs boson gives direct acess to this coupling - a very important SM 
ingredient still lacking a direct measurement.
These are some of the reasons why top quark production is a key topic  on the physics programs of the Tevatron and LHC
experiments. 

Top quarks are produced in pairs, via strong interaction, or via electroweak interactions leading to ``single top quark" final states,
either in association with jets (in the $t$-channel or $s$-channel) or with a W boson (in the tW channel). 
In the SM they decay almost exclusively into a b quark and a W boson. Other allowed decay modes are strongly suppressed. 
Decay channels of the top quarks are named according to
the decay of the W boson: either leptonic (into a lepton and a neutrino) or hadronic (into a quark-antiquark pair).

Differences on the incident hadrons and collision energy at the Tevatron and LHC colliders allow
different aspects of top quark production to be tested. At the Tevatron, top quark processes from
 p$\bar{\rm p}$ collisions at 1.96 TeV are mostly initiated by quark-quark interactions. At the LHC, 
 in pp collision at 7 and 8 TeV, gluon-initiated processes are dominant. 
Therefore, quark-initiated processes, such as $s$-channel single top quark production are
relatively more abundant at the Tevatron, while other mostly gluon-initiated processes such as the tW production, or those requiring higher
energies, such \ttbar Z and \ttbar W associated productions,
are not accessible at the Tevatron, being their  cross sections small even at the LHC s.c.m. energies. 

In this contribution, the most recent measurements on top quark production from 
the ATLAS \cite{atlas}, CDF, 
CMS \cite{cms} and D0 
 Collaborations are discussed. 
Measurements of single top quark production, inclusive \ttbar~production in nearly all decay modes, differential
\ttbar~cross sections and associated production are presented.
Some topics, such as jet multiplicity in \ttbar~and searches
for \ttbar H production, are covered in other contributions to this conference, and are not discussed here.

Along with the experimental efforts, precise NNLO calculations have been developed in the last years. In a fruitful 
 synergy between theory and experiment, data observation
leads to a finer tune of the theoretical inputs, such as the proton PDF and QCD renormalization and factorization scales, while
 experiments profit from instance from the development of more accurate Monte Carlo (MC) programs  which are used for the estimation of 
efficiencies and corrections of the data. High precision 
 both at the experimental and the theoretical sides enhances the possibility of signs of new physics to be observed in data, as deviations from the SM.

\section{Single top quark production}

Single top quark production is characterized by very small signal yields over an overwhelming background.
These competing final states come mainly from \ttbar, W+jets, Z+jets, QCD multijes and diboson events.
For this reason, analysis techniques for single top quark cross sections measurements involve 
the use of event topology for signal to background separation, in addition to
optimized event selection. Boosted decision trees (BDT), neural networks or multivariate
likelihoods are examples such techniques exploiting the full event kinematics to separate signal from background.

Associated production of a single top quark and a W boson is not visible at the Tevatron, due to very small yields. At the LHC,
measurements are performed with both W bosons (from the top decay and the prompt W) decaying leptonically, leading to a final state
with two leptons and a b jet. With only one additional b jet, the final state for \ttbar~events decaying dileptonically is found,
with a cross section more than 10 times larger.
In a very large amount of \ttbar~events, one of the b jets misses reconstruction, and the event is selected as tW candidate. 
CMS measurement \cite{twcms}, based on a sample of integrated luminosity 12.2 fb$^{-1}$ at 8 TeV, uses 
events containing exactly two leptons ($ee,e\mu,\mu\mu$) and exactly 1 b-tagged jet optimized for
the signal. The dominant \ttbar~background is estimated using a control region 
containing one additional b-tagged jet. The separation of signal and background is achieved on a BDT
analysis using 13 event variables, yielding $\sigma_{\rm tW}=23.4\pm5.4$ pb. 
The measurement corresponds to the first observation of 
single top quark production in the tW channel, with an observed significance of 6.1$\sigma$ (with 5.4$\pm 1.4\sigma$ expected). 
ATLAS measurement \cite{twatlas}, based on 20.3 fb$^{-1}$ of 8 TeV data, uses two samples, both containing an $e\mu$ pair, and either 1 or 2 jets. The sample with one jet
is used in a BDT with 19 variables for the signal estimation, while 
the 2 jets sample is used to
determine the \ttbar~normalization in a BDT with 20 variables.  The measurement, $\sigma_{\rm tW}=27.2\pm2.8$ (stat) $\pm5.4$ (syst.) pb, 
is an evidence for tW channel production at  4.2$\sigma$ (with $4\sigma$  expected).
Both measurements are in agreement with the SM predictions at approximated NNLO, of $22.2\pm0.6$ (scale) $\pm1.4$ (PDF) pb.

\begin{figure}[t]

\vspace{-1.6cm}
\hspace{1cm}\includegraphics[height=4.4cm]{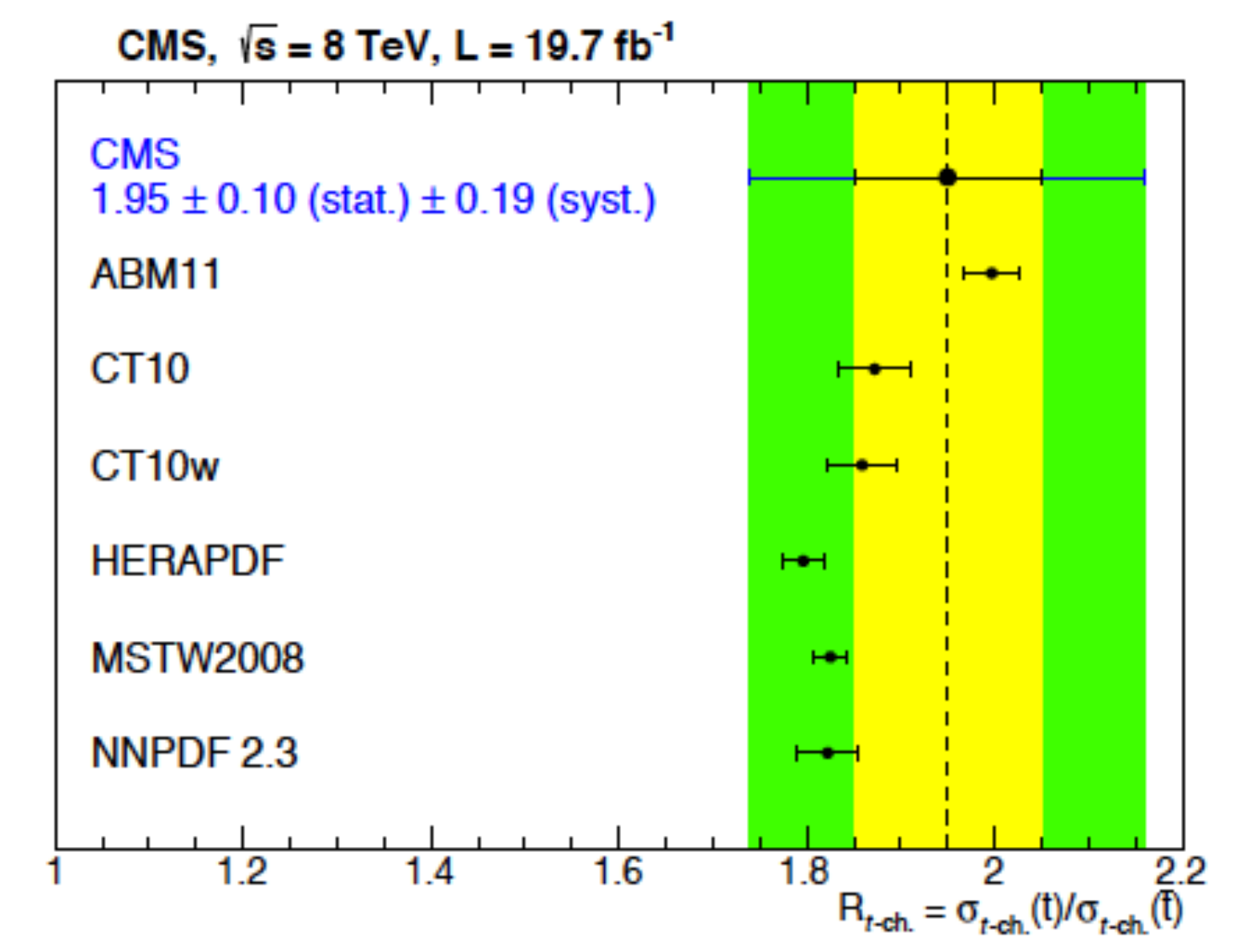} 

\vspace{-6.6cm}
\hspace{8.cm} \includegraphics[height=9.cm]{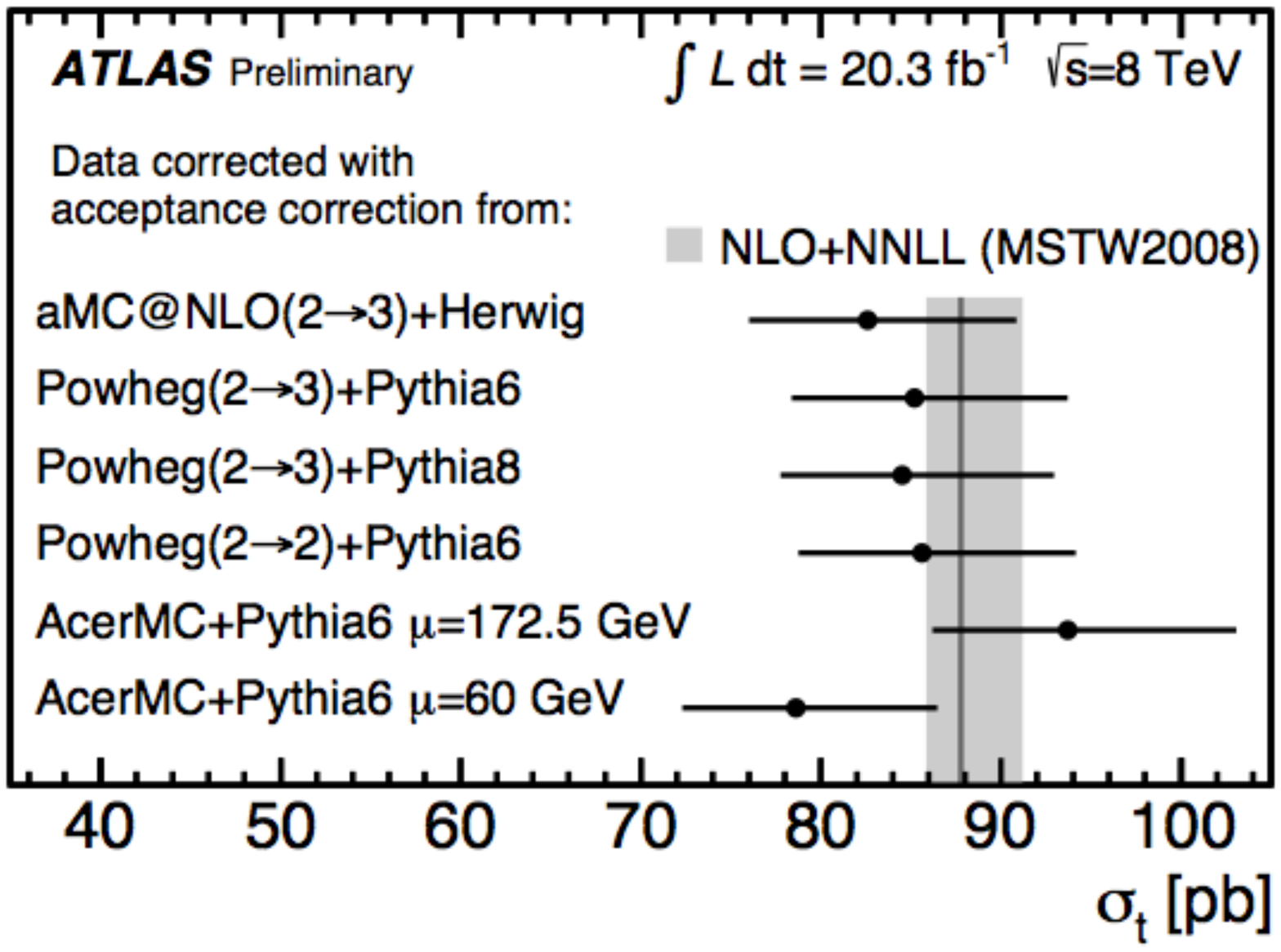}

\vspace{-2.8cm}
\caption{Studies by CMS \cite{cmsratio} (left) and ATLAS \cite{atlasratio} (right)  using single top quark events in the $t$-channel.}
\label{fig:tchanstudies}
\end{figure}

Single top quark production in the $t$-channel has been observed long ago both at the Tevatron and at the LHC.
Currently, uncertainties on the most precise cross section measurements are around 20\% at 1.96 TeV and 10\% at 7 \cite{cms7} and 8 TeV.
As more data becomes available at the LHC,  $t$-channel events can be used in more detailed studies.
For instance, the measurement of the ratio between tops and anti-tops produced in the
$t$-channel was performed by ATLAS \cite{atlasratio} and CMS \cite{cmsratio} Collaborations. Since the ratio is mainly driven by the relative
proportion of quarks up and charm (leading to t quark production) and quarks down and strange ($\bar{\rm t}$ quark production) 
in the proton,
the measurement is sensitive to the parametrizations of the proton PDF.
Fig.  \ref{fig:tchanstudies}(left) shows the results for the CMS Collaboration: the measured ratio 
is compared to SM predictions using several PDF parametrization.
ATLAS measurement yields similar results.
ATLAS Collaboration studied the dependence of the model used for acceptance correction in the measurement of
 $t$-channel cross section \cite{ATLAS-CONF-2014-007}. The  $t$-channel generator choice 
is, together with jet energy corrections uncertainty, the dominant source of uncertainty on the measurement of the
 $t$-channel fiducial cross section. Several NLO MC generators were tested, as well as the LO {\sf AcerMC} generator with different 
factorization and renormalization scales. Once the fiducial cross section is extrapolated to the full phase space,
NLO generators models are in much better agreement to each other, and other uncertainties on the extrapolation 
become dominant.
Results are presented in Fig. \ref{fig:tchanstudies}(right).
Although with  current precision no particular PDF parametrization or MC model can be excluded, 
these studies demonstrate the potential of $t$-channel events in future analyses as stringent tests of the SM, 
which may reduce systematic uncertainties on  top production measurements.

The single top-quark production mode with smallest cross section in the SM, the $s$-channel, was first measured  at the Tevatron
in combination with the $t$-channel. The typical event selection criteria was as generic as possible, 
requiring 1 lepton, at least 2 jets (from which at least one b tagged) and some  $E_{\rm T}^{\rm miss}$. 
D0 \cite{d0st} combines three multivariate techiniques, using up to 30 variables to separate signal from background.
CDF \cite{cdfst} bases its measurement on NN analyses using up to 14 variables.
A third measurement from CDF \cite{cdfmet} uses an even looser selection criteria, requiring events with
 $E_{\rm T}^{\rm miss}$, 2 or three jets, but vetoing isolated leptons. This selection recovers events where
the  lepton misses reconstruction, keeping the sample orthogonal to that in the previous measurement.
The precision on these measurements range from 14 to 48\%.

In addition to the difficult task of separating single top quark $s+t$-channel events from the background, Tevatron experiments
used properties of event topology to separate $s$- and $t$-channel events from each other. Events in the $t$-channel 
tend to have a distinctive forward jet, whose direction in p$\bar{\rm p}$ collisions is correlated with the lepton charge, 
and one b jet. On the other hand, 
$s$-channel events are more likely to have central jets, two of them from b quarks. 
Using these properties, CDF and D0 \cite{cdfd0s} 
have formed multivariate discriminants optimized for $s$- and $t$-channel
 separately. 
These discriminants were used on  a Bayesian statistical analysis to combine the results from both collaborations.
The three individual
measurements above mentioned were combined, yielding the only existing observation of single top quark production 
in hadron colliders to this date: 
$\sigma_s= 1.29^{+0.26}_{-0.24}$ at 6.3 standard deviations (with 5.1 expected).

With relatively less quark-initiated processes as compared to the Tevatron, LHC has not yet observed 
the production of  $s$-channel single top quark. 
With a cut-and-count analysis based on 0.7 pb$^{ -1}$ of 7 TeV data, 
ATLAS \cite{atlassch} has
set an upper limit of 26.5 fb at 95\% CL to the production. 
CMS Collaboration \cite{cmssch} has performed a more sophisticated BDT analysis 
based on 19.3 fb$^{ -1}$ of data at 8 TeV, measuring $\sigma_s= 6.2\pm 5.4\pm 5.9$ pb with 0.7 $\sigma$ (with 0.9 expected). 

\section{\ttbar~production cross section}

The production of top quark pairs is an excellent ground to explore the frontiers of the SM.
It provides important inputs to improve current limitations of the
SM predictions and also plays an important role on direct searches for physics beyond the SM. 
Signs of new physics could be evidenced as an enhancement of \ttbar~production in a particular decay mode,
or have SM \ttbar~as background.

\begin{figure}[htb]
\includegraphics[height=4.5cm]{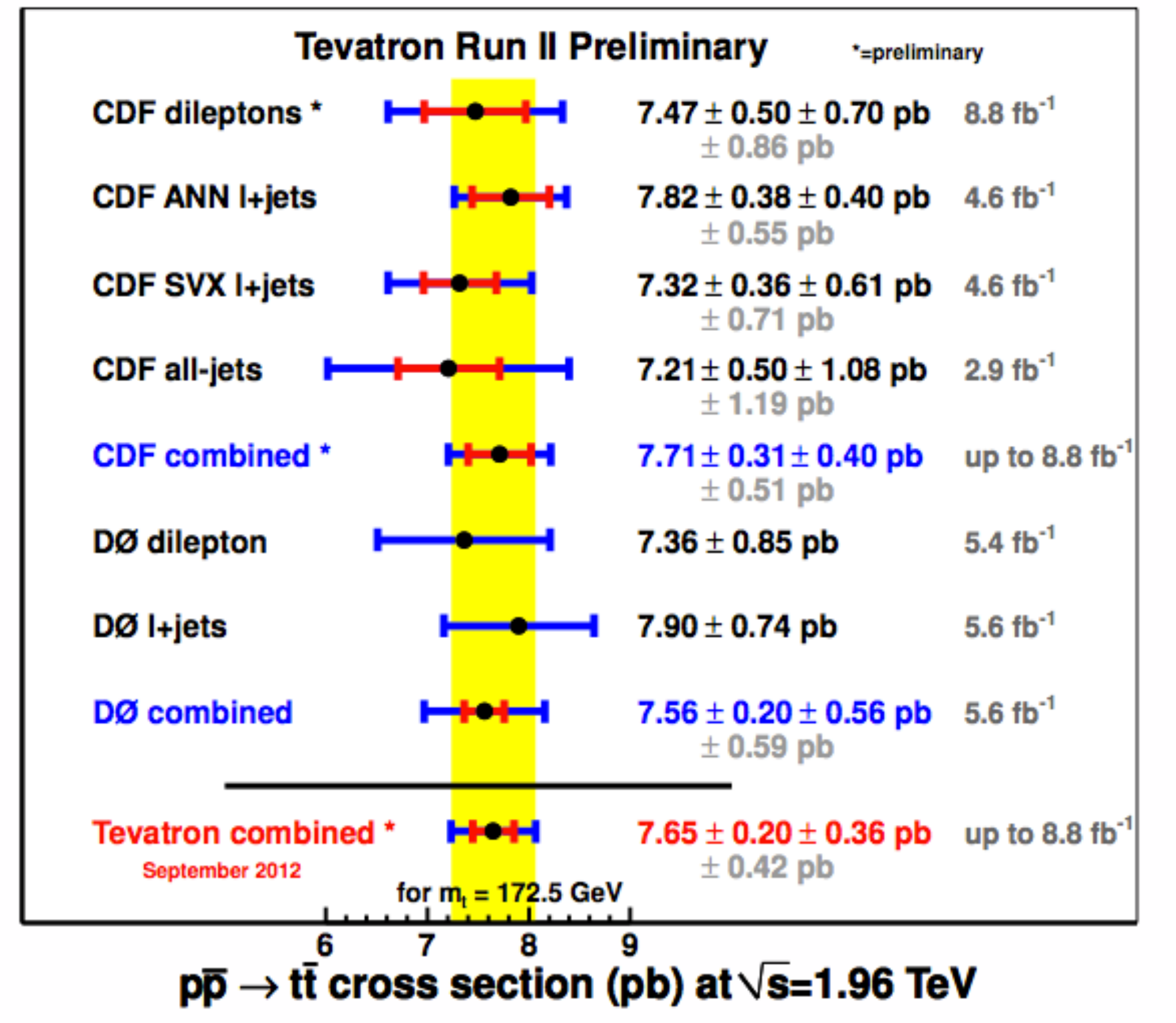}

\vspace{-5cm}
\hspace{6.5cm}\includegraphics[height=5cm]{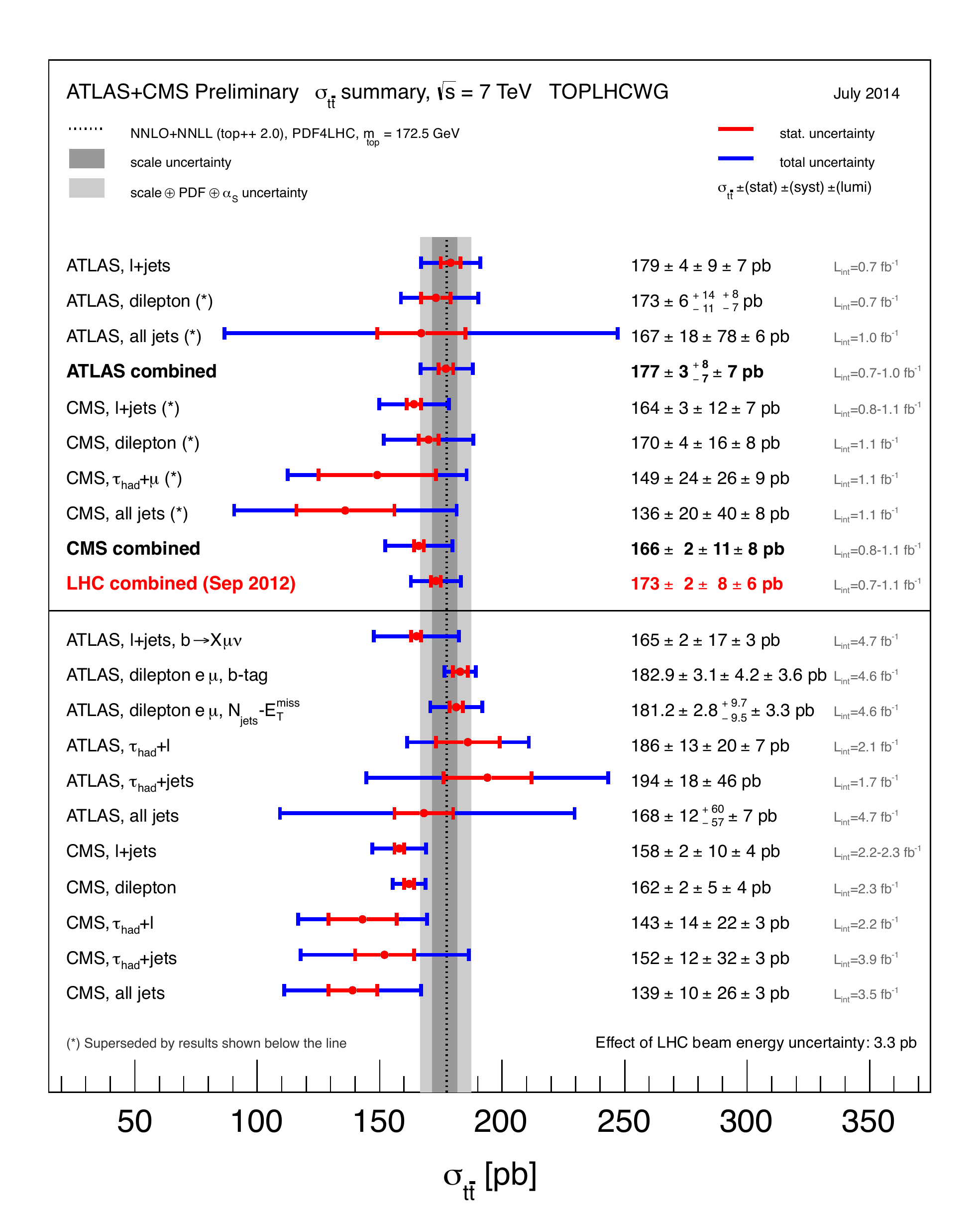}

\vspace{-5cm}
\hspace{11.5cm}\includegraphics[height=5cm]{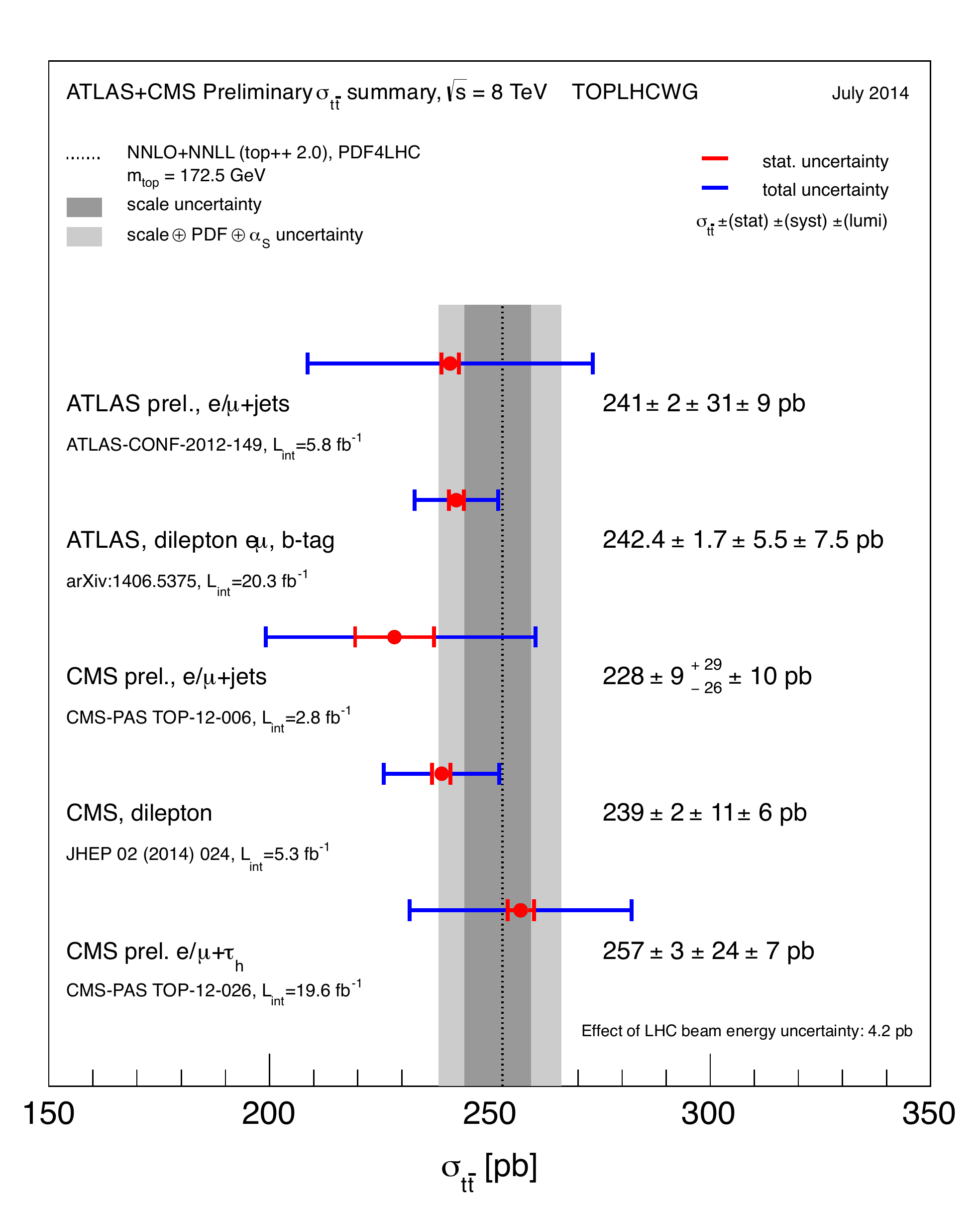}

\vspace{-0.5cm}
\caption{Summary of top cross section measurements at the Tevatron \cite{d0cdfttcomb} and LHC \cite{summaryplot}.}
\label{fig:figure1}
\end{figure}

The total \ttbar~inclusive cross section has been measured both at the Tevatron and at the LHC, at the
centre-of-mass energies of 1.96, 7 and 8 TeV. 
At the Tevatron, CDF and D0 Collaborations \cite{d0cdfttcomb} reported on a combination of 4 
measurements from CDF Collaboration and 2  from D0 Collaboration, substantially improving the precision
on the individual measurements.
Systematic uncertainties related to the modeling of signal are the largest, and theoretical uncertanties on
the background are also important.  
The most precise measurements at 8 TeV are those using fullleptonic \ttbar~decays. In a cut-and-count analysis
using ee,e$\mu$ and $\mu\mu$ and 2 b-tagged jet final states, and based on integrated luminosity of 5.3 fb$^{-1}$, 
CMS \cite{CMSdilep} measures the cross section with a precision of about 4.5\%. ATLAS \cite{ATLASemu} uses a different cut-and-count analysis,
where
by separating events with 1 and 2 b-tagged jets, the efficiency of reconstructing and tagging a b jet can be extracted
from the data simultaneously with the cross section, further reducing  systematic uncertainties. 
ATLAS measurement is based on e$\mu$ final states and 20 fb$^{-1}$.
A summary of the most recent measurements, compared
to the latest theoretical predictions, is shown in Figure \ref{fig:figure1} and Table \ref{tab:table1}. 
All possible decay channels, except that where both W bosons decay to taus, have been analyzed 
and found in very good agreement with the SM predictions. It is interesting to notice that theory and experiment have reached 
a similar level of precision, being the dominant experimental uncertainties in most cases precisely those arising from
theoretical uncertainties on the signal modeling.

\begin{figure}[htb]
~ \vspace{-2cm}

\includegraphics[height=3.5in]{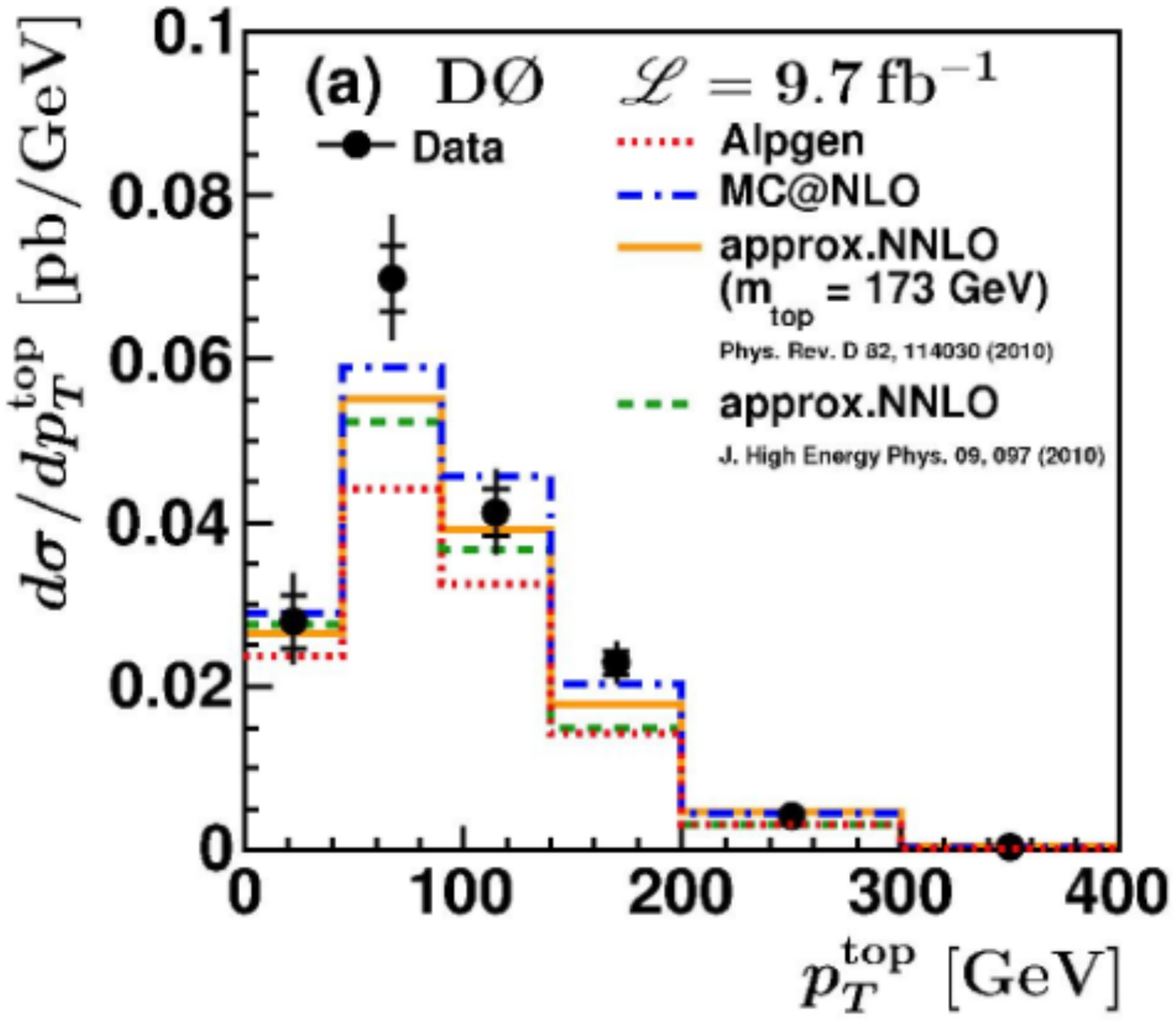} 

\vspace{-6.8cm}
\hspace{6cm} \includegraphics[height=1.8in]{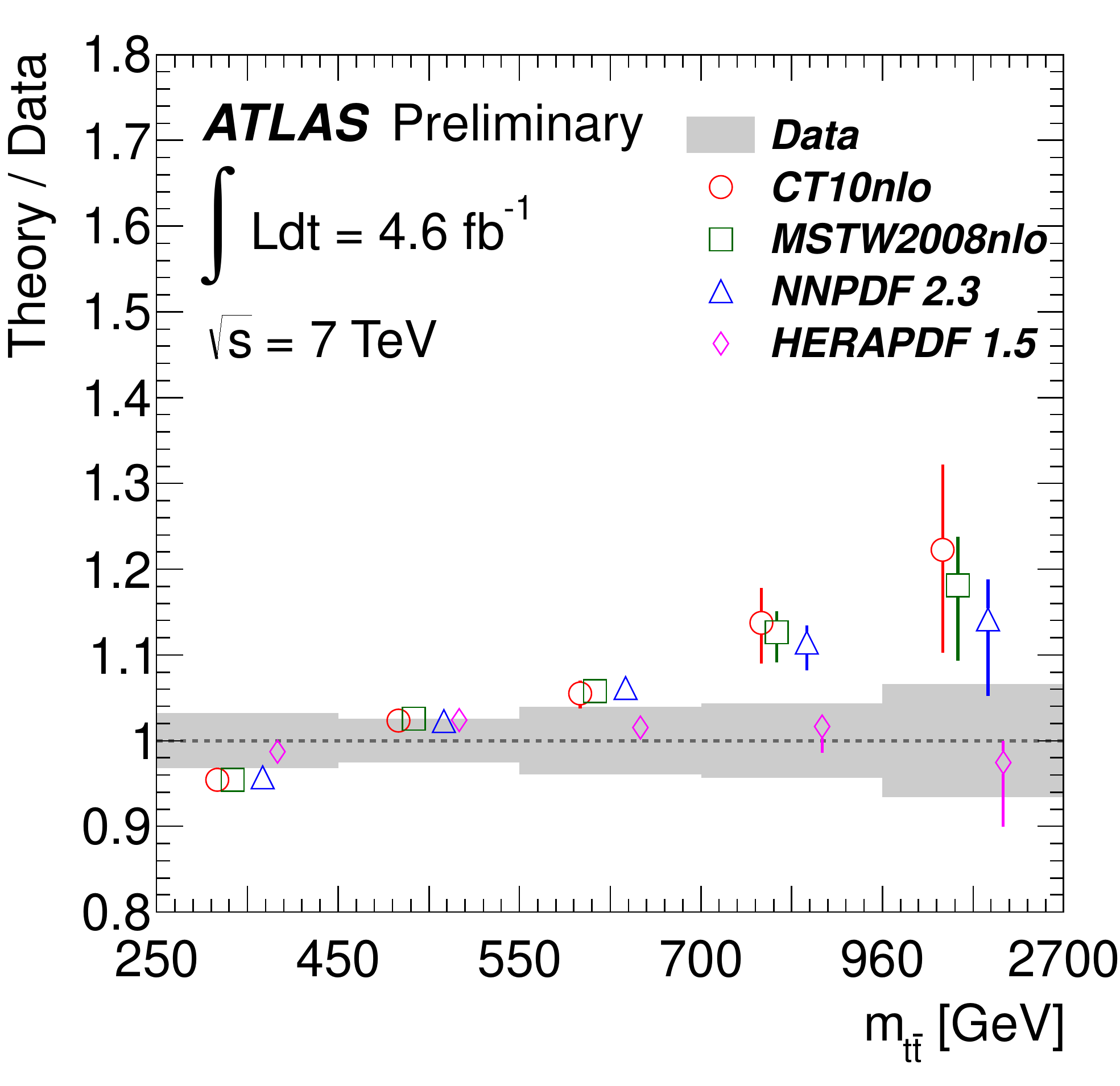} 

\vspace{-8cm}
\hspace{9.5cm} \includegraphics[width=8cm]{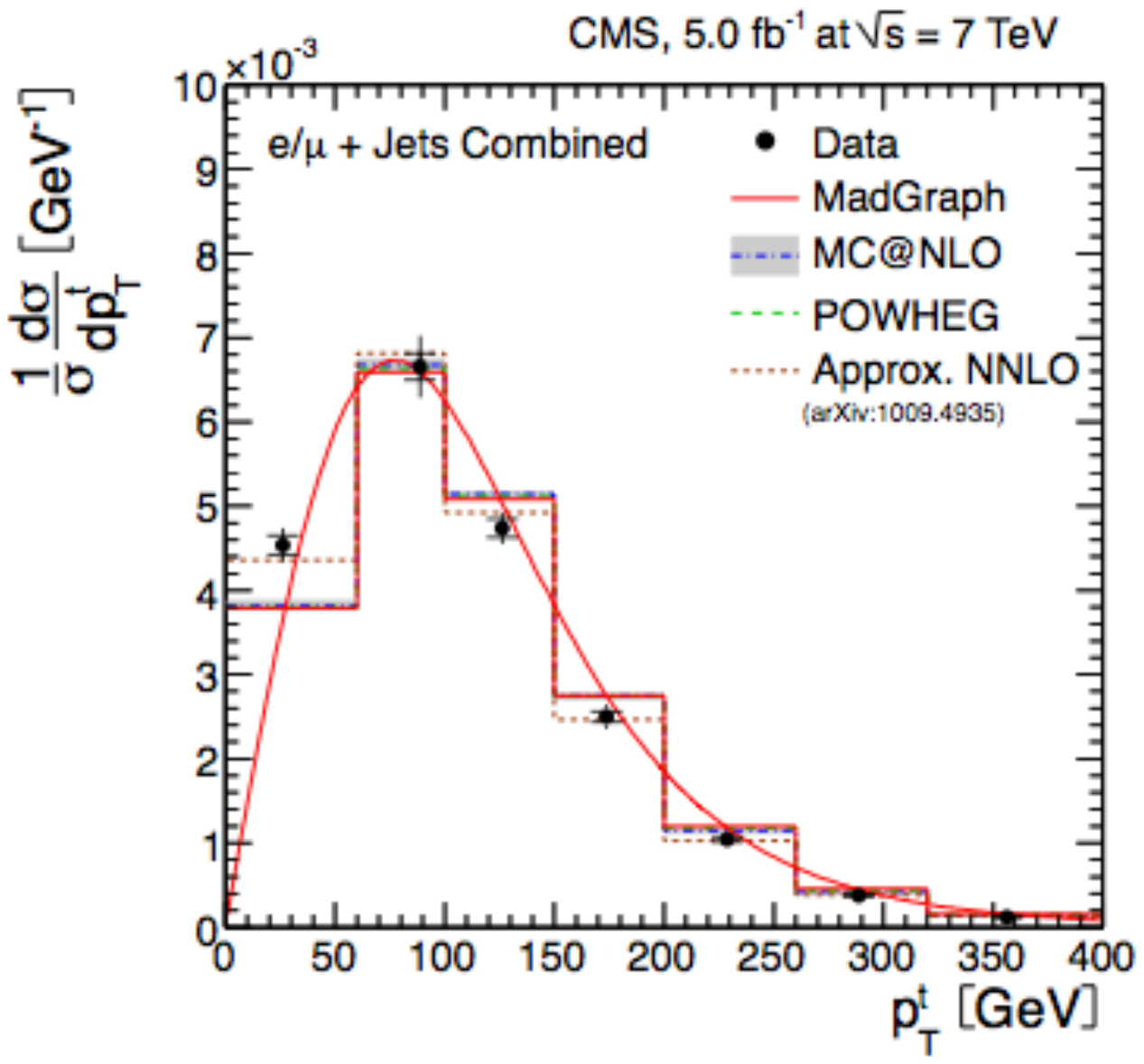}

\vspace{-3.9cm}
\caption{Differential \ttbar~cross sections distributions from D0 \cite{d0diff} (left) ATLAS \cite{atlasdiff} (center) and CMS \cite{cmsdiff} (right).}
\label{fig:diff}
\end{figure}

Recently, a new realm has started where top quark production can be investigated with unprecedented 
level of scrutiny: the measurements of differential \ttbar~cross section
 as a function of several kinematic variables. 
D0 \cite{d0diff}, ATLAS \cite{atlasdiff} and CMS \cite{cmsdiff} 
collaborations have measured differential cross sections at 1.96, 7 and 8 TeV, normalized to
their respective total inclusive cross sections. The total uncertainty 
for the D0 measurement,  using  the lepton+jets decay modes,  is dominated by statistical uncertainties
, while typical systematic uncertainties range between 8\% and 23\%. 
At the LHC, where both lepton+jets and full leptonic channels were used similar values ($\sim 10$ to 20\%) account for the total
uncertainty.
Even with the excellent agreement observed between data and theory, these high-precision measurements are able to reveal 
finer details of the models that still have some ground for improvement. 
Some examples are given in Fig. \ref{fig:diff}:  MC@NLO predictions are in better agreement with D0 data than Alpgen;
ATLAS has shown the dependency of the several PDF parametrizations as a function of the mass of the 
\ttbar~system;
the description of the top quark $p_{\rm T}$ spectra by the current MC models used by the 
LHC Collaborations was found to have small deviations from the data, as observed by both ATLAS and CMS 
independently of the decay mode and collision energy.  
Approximate NNLO and NLO+NNLL predictions, when available, 
are able to describe well the data.

\begin{table}[t]
\begin{center}
\begin{tabular}{l|c|l|c} \hline 
Proc. & $\sqrt{s}$ & Measurement (pb) &  Theory prediction (pb) \\ \hline
 \ttbar & 1.96  &  CDF \& D0 comb. \cite{d0cdfttcomb}: $7.6\pm0.2\pm0.36$   & NNLO+NNLL \cite{tevttTheory}:$7.35^{+0.17}_{-0.12}{\rm (PDF)}^{+0.11}_{-0.21}{\rm (QCD\ scale)}$  \\ \cline{2-4}  
 & 8 & ATLAS \cite{ATLASemu}: $237.7\pm1.7\pm7.4\pm7.4\pm 4.0 $  &  Full NNLO: \\ 
  & & CMS \cite{CMSdilep}: $239.0\pm2.1\pm11.3\pm 6.2$ &  $252.9\pm11.7{\rm (PDF \& \alpha_s)}^{+6.4}_{-8.6}{\rm (QCD\ scale)}$ \\ \hline
\ttbar$\gamma$ & 1.96  & CDF \cite{cdfttgamma}:  0.18$\pm$0.07$\pm$0.04$\pm$0.01  & LO + K factor \cite{cdfttgammatheory}: 0.17$\pm$0.03   \\ \cline{2-4}
  & 7 & ATLAS $p_{\rm T}^\gamma>$8 GeV \cite{atlasttgamma}:  $2.0\pm0.5\pm0.7\pm0.08$  &  LO + K factor  \cite{atlasttgammatheory}:  $1.8\pm 0.5$   \\ \cline{2-4}
  & 8  & CMS $p_{\rm T}^\gamma>$20 GeV \cite{cmsttgamma}:  $2.4\pm0.2\pm0.6$   &   LO + K factor \cite{atlasttgammatheory}: $2.1\pm 0.4$ \\ \hline    
\ttbar Z & 8 &  CMS \cite{cmsttz}: $200^{+80}_{-70}$$^{+40}_{-30}$ fb  & NLO \cite{ttztheory}: $197^{+22}_{-25}$ fb \\ \cline{2-4} 
\ttbar W & 8  & CMS \cite{cmsttz}: $170^{+90}_{-80}$$^{+70}_{-70}$  fb  & NLO \cite{ttztheory}: $206^{+21}_{-23}$ fb\\ 
\hline
\end{tabular}
\caption{Summary of the most precise measurements of  \ttbar(+ $\gamma$,V) production. $\sqrt{s}$ are displayed in TeV and 
cross sections in pb or fb. Experimental
uncertainties are displayed as $\pm$stat.$\pm$syst.$\pm$lumi. ATLAS  \ttbar~cross section at 8 TeV quotes an additional uncertainty
due to the uncertainty on the LHC beam energy.}
\label{tab:table1}
\end{center}
\end{table}

\section{Associated production: \ttbar Z, \ttbar W, \ttbar $\gamma$}

Associated production of top pairs with electroweak gauge bosons is one of the most  
interesting subjects in top quark production, since it can immediately be used to test SM predictions
 and in the near future provide important information to improve
electroweak measurements at the LHC (and even to the next generation of particle colliders).
In particular, \ttbar$\gamma$ cross section can be reinterpreted in terms of the top quark electrical charge - 
therefore its measurement can be considered and alternative probe of that physical quantity.
The  \ttbar Z production cross section is proportional to the  \ttbar Z electroweak gauge coupling,
and can be use to impose stringent limits 
on anomalous top quark couplings, with severe implications on many new physics models. 
And all of them, together with \ttbar W, are part of the irreducible 
background on the measurement of \ttbar H production, and share the same final state with many 
processes beyond the SM.

 Among these three associated production processes, only \ttbar$\gamma$ is  accessible at the Tevatron. 
The dominant background to this process is instrumental, when a QCD jet satisfies all selection criteria normally passed by
a photon. 
It has been measured by the CDF Collaboration  \cite{cdfttgamma} using semileptonic \ttbar~decays on a selected sample of 30 events. The  main background
is estimated 
using jets passing most of photon selection criteria, but failing at least one of them.
At the LHC, ATLAS \cite{atlasttgamma} and CMS measurements \cite{cmsttgamma} at 7 and 8 TeV, respectively, use template fits on variables with large discriminating
power to separate real from fake photons.
Results are presented in Table \ref{tab:table1}, in comparison to the predictions from the SM.

Measurements of  \ttbar V typically use leptonic decays of top quarks and V bosons,
leading to final states containing 2 to 4 leptons, 
with characteristic lepton charge combination for each process.
Given that SM processes with multilepton final states are rare, backgrounds to these processes are also mostly instrumental,
 from QCD jets misidentified as leptons or misidentified lepton charge. 

CMS measurement \cite{cmsttz}, presented at this Conference for the first time, 
is based on 19.5 fb$^{-1}$ of data at 8 TeV, and uses events with three final states containing electrons and/or muons: 
2 same-sign leptons, aiming to tag semileptonic \ttbar + W boson events, and 3 or 4 leptons to identify \ttbar Z events. 
The instrumental background is estimated via data-driven techniques: jet-enriched samples used to 
control the rate of fakes and charge misassignment is studied on opposite sign ee or e$\mu$ events, predominantly from Drell-Yan or 
\ttbar~processes. Other backgrounds are estimated on MC. All final states are combined
on a profile likelihood analysis.
A simultaneous fit to all channels yields a combined cross section $\sigma_{\rm t\bar{t} V}=380^{+100}_{-90}$(stat.)$^{+100}_{-90}$(syst) fb 
with 3.7 $\sigma$ significance.
 Then, from fits using same-sign (the remaining) channels  \ttbar W (\ttbar Z) cross section is extracted. Two-dimensional fits where
\ttbar W and \ttbar Z are extracted simultaneously yield the same results.
Results for the individual channels are given in Table \ref{tab:table1}.

\section{Summary and conclusions}

The {\it era of observations and evidences} of the top quark production was initiated at the Tevatron collider in 1995, when a few dozen of \ttbar~events were 
observed for the first time. Tevatron also holds the first evidence for single top quark production, and its $s$-channel observation. 
At the LHC, single top quark production was observed at the $t$-channel and tW channels. Most recently, \ttbar~processes in association to bosons, at much lower
cross sections, have also been observed.

By the time Tevatron ceased operations, CDF and D0 collaborations had each collected samples of almost 70.000 top pairs. 
Meanwhile, LHC ramped up activities as a ``top factory", collecting more than 5.5 million top pairs and
2.7 million single top events per experiment available for detailed studies. 
 These millions of top quark events have brought us to the {\it precision era} today, where
 theory and experiment challenge each other to reach better and better accuracy.

Measurements of total inclusive cross sections of top quark pairs have reached an experimental precision of about 5\% 
at the three  probed s.c.m. energies (1.96, 7 and 8 TeV), while single top quark processes have been measured with
precisions of 9-30\%, depending on the production channel and collision energy. 
This accuracy is mainly limited by the uncertainties on the modelling of the top signal MC. For single top
quarks, the statistical precision is also important, specially in the measurements using Tevatron data.


Differential cross sections have the potential to constrain some theoretical uncertainties in the signal modelling. 
For instance, with current precision, differential cross sections data, such as jet multiplicity, 
 can rule out extreme variations of the QCD factorization and renormalization scales; 
ratios of single top quark-to-antiquark production in the t-channel is sensitive to the choice of the proton PDF, 
although with the current precision no set of PDFs is favoured.
Reducing theoretical uncertainties, the dominant systematic uncertainties in most of the top quark measurements, 
is a prospect for improving future measurements of top quark production. 

The associated production of top pairs with photons, Z's and W bosons have also been measured. These production processes
at extremelly low cross sections have finally been observed, yielding significant tests of the SM: firstly, 
the cross sections proportional the magnitude of the electroweak couplings to the top quark are very sensitive to
new physics; secondly they are an important milestone on the measurement of the SM \ttbar H production.
Since these processes are statistically limited, their precision at the planned LHC high luminosity runs at 
13 TeV is expected to improve. 
The precision of the direct measurements of the top-Higgs Yukawa coupling achieved at the LHC in the next years
may have a deep impact on the future generations of colliders, since they are, along with the discovery or
further investigation of 
new physics, at the core of physics goals for planned experiments such as the International Linear Collider. 

So far, no sign of new physics have been seen in top quark production measurements at 1.96, 7 and 8 TeV.
These processes will continue to be tested on a new energy regime when the LHC resumes operation in 2015.
If new physics exists 
at an scale reachable at the LHC, and couples to the top quark, it may become evident on deviations
between the very high precision data and theory. In this case,  the {\it era of new physics discovery} may be started soon.


\end{document}

%% file: econfmacros.tex
\def\beq{\begin{equation}}
\def\eeq#1{\label{#1}\end{equation}}
\def\eeqn{\end{equation}}

\def\beqa{\begin{eqnarray}}
\def\eeqa#1{\label{#1}\end{eqnarray}}
\def\eeqan{\end{eqnarray}}

\let\bar=\overbar

\def\Dslash{\not{\hbox{\kern-4pt $D$}}}
\def\dslash{\not{\hbox{\kern-2pt $\del$}}}

\def\msb{{\bar{\ssstyle M \kern -1pt S}}}

